\documentclass{article}
\usepackage{amssymb}

\def\l{{\lambda}}

\def\on#1#2{\mathop{\vbox{\ialign{##\crcr\noalign{\kern2pt}
$\scriptstyle{#2}$\crcr\noalign{\kern2pt\nointerlineskip}
\kern-2pt$\hfil\displaystyle{#1}\hfil$\crcr}}}\limits}
\def\nn{ \nonumber }
\def\bq{ \begin{equation} }
\def\eq{ \end{equation} }
\def\ben{ \begin{eqnarray} }
\def\en{ \end{eqnarray} }
\def\ll{ \label }
\def\frac#1#2{{#1\over #2}}
\def\dfrac#1#2{{\displaystyle{#1\over#2}}}

\begin{document}

\title{Canonical transformations of the time
for the Toda lattice and the Holt system.}
\author{
 A.V. Tsiganov\\
{\small\it
 Department of Mathematical and Computational Physics,
 Institute of Physics,}\\
{\small\it St.Petersburg University, 198 904,  St.Petersburg,
Russia,}\\ {\small\it e-mail: tsiganov@mph.phys.spbu.ru} }
\date{}
\maketitle

\vskip0.5cm
{ For the Toda lattice and the Holt system we consider properties
of canonical transformations of the extended phase space, which
preserve integrability. The separated variables are invariant
under change of the time. On the other hand, mapping of the time
induces transformations of the action-angles variables and a shift
of the generating function of the B\"{a}cklund transformation.}

\section{Introduction.}
\setcounter{equation}{0}
Let $\cal M$ be  a $2n$-dimensional symplectic manifold  (phase
space) with coordinates $\{p_j,q_j\}_{j=1}^n$. The Hamilton
function $H(p,q)$ defines the hamiltonian dynamical system on
$\cal M$. Here $p$ and $q$ denote $p_1,\ldots,p_n$ and
$q_1,\ldots,q_n$, respectively.

By adding to $\cal M$ the time $q_{n+1}=t$ and the Hamiltonian
$p_{n+1}=-H$ one gets $2n+2$-dimensional extended phase space
${\cal M}_E$  of the given hamiltonian system \cite{lanc49}.
Canonical functional $S$ on ${\cal M}_E$ has the following
completely symmetric form
\[S=\int_{\tau_1}^{\tau_2}\sum_{i=1}^{n+1} p_i\,q'_i\,d\tau\,.\]
On the extended phase space ${\cal M}_E$ the Jacobi,
Euler-Lagrange and Hamilton variational principles $\delta S=0$
are differed by an additional constraint
\bq
{\cal H}(p_1,\ldots,p_{n+1};q_1\,\ldots,q_{n+1})=0\,.\label{oham}
\eq
Here $\cal H$ is called generalised Hamilton function
\cite{lanc49}, which determines the evolution.

By definition the Hamilton function $H(p,q)$ is a function on
$\cal M$. At the same time the Hamiltonian $H$ and the time $t$
are variables in ${\cal M}_E$, which are independent on the other
variables $(p,q)$. Of course, equation of the zero-valued energy
surface ${\cal H}=0$ may be rewritten as $H(p,q)=H$ \cite{lanc49}.
Thus, unless other wise indicated, $H(p,q)$ denotes a function on
$\cal M$, and $H$ denotes independent variable in ${\cal M}_E$.

By definition canonical transformations of the extended phase
space ${\cal M}_E$ preserve the Ha\-mil\-ton-\-Jacobi equation and
differential form
\[\alpha=\sum_{j=1}^{n} p_j\,dq_j-Hdt\,.\]
So, any canonical transformation of the time looks like
\[  d\widetilde{t}=v^{-1}(p,q)\,dt
\qquad \widetilde{H}={v(p,q)}\,H\,,
\]
where we used implicit transformation of the time $t$ as in the
general relativity.

Any canonical transformation of the initial phase space $\cal M$
maps any integrable system into the other integrable system.
However, we have not a regular way to obtain canonical
transformation of the extended phase space ${\cal M}_E$, which
maps a given integrable system into the other integrable system.
Nevertheless, we can try to construct such transformations by
using different approaches developed for the integrable system.

In \cite{ts98b,ts99c} some canonical transformations of the time
have been constructed for the Toda lattices and for the
St\"{a}ckel systems by using transformations of the Lax matrices.

The aim of this letter is to study some properties of the such
canonical transformations of the extended phase space. For the
Toda lattice and the Henon-Heiles system we shall prove that
separated variables are invariant by the change of the time. On
the other hand, mapping of the time induces transformations of the
action-angles variables and a shift of the generating function of
the B\"{a}cklund transformation.

\section{The Toda lattice.}
\setcounter{equation}{0}
The periodical Toda lattice is described by the following Hamilton
function
\bq
H(p,q)=
\dfrac12\,\sum_{i=1}^n \,p_i^2+a_i\,e^{q_i-q_{i+1}}\,.\label{anh}
\eq
Here $\{p_i,q_i\}$ are canonical variables and the periodicity
conventions $q_{i+n}=q_i$ and $p_{i+n}=p_i$ are always assumed for
the indices of $q_i$ and $p_i$.

The $n\times n$ Lax matrices \cite{km75,fml76a} for the Toda
lattice are
\ben
{\cal L}^{(n)}(\mu)=\sum_{i=1}^n
p_j\,E_{i,i}&+&\sum_{i=1}^{n-1}\left(\,
e^{q_i-q_{i+1}}\,E_{i+1,i}+E_{i,i+1}\right)+\nn\\ &+&
\mu\,e^{q_n-q_1}\,E_{1,n}+\mu^{-1}\,E_{n,1}\,,\label{alax}
\en
\[
{\cal A}^{(n)}(\mu,q)=\sum_{i=1}^{n-1}\,
e^{q_i-q_{i+1}}\,E_{i+1,i}+
\mu\,e^{q_n-q_1}\,E_{1,n}
\]
where $E_{i,k}$ stands for the $n\times n$ matrix with unity on the
intersection of the $i$th row and the $k$th column as the only
nonzero entry.

By abuse of notation we shall omit the superscript $n$ of the Lax
matrices (\ref{alax}) when it is not important. The exact solution
of the equations of motion is due to existence of the Lax
representation \cite{km75,fml76a}
\[\{H(p,q),{\cal L}\}=[{\cal L},{\cal A}]\,.\]

Another $2\times 2$ Lax representation \cite{ga83,skl85a} for the
same Toda lattice is equal to
\bq
T^{(1\ldots n)}(\l)=L_1(\l)\cdots L_n(\l)=
\left(\begin{array}{cc}
  {\mathbb A}(\l) & {\mathbb B}(\l) \\
  {\mathbb C}(\l) & {\mathbb D}(\l)
\end{array}\right)
\,,
\label{22toda}
\eq
 where
 \bq
 \qquad L_j=\left(\begin{array}{cc}
 \l+p_j &\, e^{q_j} \\
 -e^{-q_j}& 0
\end{array}\right)\,,\qquad
A_j=\left(\begin{array}{cc}
 \l &e^{q_j} \\
 -e^{-q_j}& 0
\end{array}\right)
\,,
\label{2alax}
\eq
such that
\[
\{H(p,q),L_j\}=L_j\,A_j-A_{j-1}\,L_j\,,
\qquad \{H(p,q),T^{(1\ldots n)}\}=[T^{(1\ldots n)}\,,A_n\,]
\,,\]
Sometimes, we shall omit the superscripts $(1 \ldots n)$ of the
monodromy matrix $T(\l)$ (\ref{22toda}), too.

According to \cite{ts99c}, canonical transformations of the
extended phase space ${\cal M}_E$
\bq
d\widetilde{t}_j=e^{q_j-q_{j+1}}\,dt\,,\qquad
\widetilde{H}_j=e^{q_{j+1}-q_{j}}\,(H+b)\,,\qquad b\in {\mathbb R}
\label{trtan}
\eq
preserve integrability for any index $1 \leq j\leq n$.

Associated with the different indexes $j$ canonical mappings
(\ref{trtan}) are related with each other by canonical
transformations of the other variables $(p,q)$. Hence, we shall
omit subscript $j$ for the new time $\widetilde{t}$ and the new
Hamiltonian $\widetilde{H}$.

Any such change of the time gives rise to the following
transformation of the Lax matrices
\bq
\widetilde{\cal L}={\cal L}-\widetilde{H}\,
E_{j,j+1}\,,\qquad\widetilde{\cal A}(\mu,q)=v^{-1}(q)\,{\cal
A}(\mu,q)\,.
\label{antodal}
\eq
and
\ben
\widetilde{T}^{(1\ldots n)}&=&
T^{(1\ldots n)}+ T^{(1\ldots j-1)}
\left(\begin{array}{cc}H+b & 0 \\ 0 & 0\end{array}\right)
T^{(j+2\ldots n)}
\nn\\
\nn\\
&=& L_1\cdots L_{j-1}\cdot\left[L_j\,L_{j+1}+
\left(\begin{array}{cc}H+b & 0 \\ 0 & 0\end{array}\right)\,\right]
\cdot L_{j+2}\cdots L_n
\,,\nn\\
\nn\\
\widetilde{A}_n(\l,q)&=&v^{-1}(q)\,A_n(\l,q)\,.\label{tr2l}
\en
Change of the time (\ref{trtan}) maps the Toda lattice into the other
integrable system. Coefficients of the polynomials
\[P(\l)={\rm tr}\,T(\l)\,,\quad
\mbox{\rm and}\quad
  \widetilde{P}(\l)={\rm tr}\,\widetilde{T}(\l)\]
are generating functions of the integrals of motion in the involution
providing complete integrability of the systems.

The corresponding transformation of the spectral curves ${\rm
det}(\,{\cal L}(\mu)+\l\,I)=0$ or ${\rm det}(\,T(\l)+\mu\,I)=0$
looks like
\bq\begin{array}{ll}
 {C}: \qquad
-{\mu}-\dfrac1{\mu} & =\l^n+\l^{n-1}\,p+
\l^{n-2}\,\left(\dfrac{p^2}2-H\right)+\ldots \\
 \widetilde{C}:\qquad -{\mu}-
\dfrac{1-\widetilde{H}}{\mu} & =\l^n+\l^{n-1}\,p+
\l^{n-2}\,\left(\dfrac{p^2}2+b\,\right)
+\ldots\,.
\end{array}
\label{todac}
\eq
Here $p=\sum p_j$ is a total momentum, $H$ and $\widetilde{H}$ are
the corresponding Hamilton functions.

The Poisson brackets relations for the $n\times n$ Lax matrices can
be expressed in the $r$-matrix form
\[
\{\,{\on{\cal L}{1}}(\mu)\,,\,{\on{\cal L}{2}}(\nu)\,\}=
[\,r_{12}(\mu,\nu)\,,\,{\on{\cal L}{1}}(\mu)]+
[r_{21}(\mu,\nu)\,{\on{\cal L}{2}}(\nu)\,]\,.
\]
Here we used the standard notations
\[{\on{\cal L}{1}}(\mu)= {\cal L}(\mu)\otimes I\,,
\qquad {\on{\cal L}{2}}(\nu)=I\otimes
{\cal L}(\nu)\,,
\]
\[r_{21}(\mu,\nu)=-\Pi\, r_{12}(\nu,\mu)\,\Pi\,,\]
and $\Pi$ is the permutation operator in ${\mathbb C}^n\times{\mathbb
C}^n$ \cite{ft87}. Change of the time (\ref{trtan}) maps the constant
$r$-matrix for the Toda lattice
\[r_{12}(\mu,\nu)=r_{12}^{const}(\mu,\nu)=
\dfrac{1}{\mu-\nu}\,\left(\,\nu\sum_{m\geq
i}+\mu\sum_{m<i}\,\right)\, E_{im}\otimes E_{mi}\] into the following
dynamical $r$-matrix
\[r_{12}(\mu,\nu)=r_{12}^{const}(\mu,\nu)+r_{12}^{dyn}(\mu,\nu)\,,\qquad
r_{12}^{dyn}(\mu,\nu)=\widetilde{\cal A}(\nu,q)\otimes
E_{j,j+1}\,,
\]
where the second Lax matrix  $\widetilde{\cal A}(\nu,q)$ and,
therefore, dynamical $r$-matrix depend on coordinates only.

The $2\times 2$ monodromy matrix $T(\l)$ (\ref{22toda}) satisfies the
following Sklyanin $r$-matrix relations
\bq
\{\,{\on{T}{1}}(\l)\,,\,{\on{T}{2}}(\nu)\,\}=
[\,R(\l-\nu)\,,\,{\on{T}{1}}(u)\,{\on{T}{2}}(\nu)\,]\,,\qquad
R(\l-\nu)=\dfrac{\Pi}{\l-\nu}\,.
\label{sklbr}
\eq
Change of the time (\ref{trtan}) transforms these quadratic
relations into the following poly-linear ones
\ben
\{\,{\on{\widetilde{T}}{1}}(\l)\,,\,{\on{\widetilde{T}}{2}}(\nu)\,\}&=&
[\,R(\l-\nu)\,,\,{\on{\widetilde{T}}{1}}(\l)\,
{\on{\widetilde{T}}{2}}(\nu)\,]\nn\\ &+&
[\,r_{12}^{dyn}(\l,\nu)\,,\,{\on{\widetilde{T}}{1}}(\l)\,]+
[\,r_{21}^{dyn}(\l,\nu)\,,\,{\on{\widetilde{T}}{2}}(\nu)\,]
\,.\nn
\en
The corresponding dynamical $r$-matrix is given by
\[
r_{12}^{dyn}(\l,\nu)=A_n(\l,q)\otimes\,\left(\, L_1\cdots
L_{j-1}\cdot \left(\begin{array}{cc}
 1 & 0 \\
 0 & 0
\end{array}\right)
\cdot L_{j+1}\otimes L_n\,\right)\,.
\]
Here all the matrices $L_k$ depend on the spectral parameter $\nu$
and $A_n(\l,q)$ is the second Lax matrix (\ref{22toda}).

Now let us look at the separated variables and the action-angle
variables in framework of the traditional consideration of the
Toda lattice. Complete list of references can be found in
\cite{km75,fml76a,skl85a,pe91}. Let us study the Toda system and the dual
system simultaneously and, for the simplicity,  consider change of
the time (\ref{trtan}) at $j=1$ such that
\[\widetilde{H}=\exp(q_2-q_1)\,(H+b)\,\]
and
\bq\widetilde{\cal L}^{(n)}=\left(
\begin{array}{cccc}
  p_1 & e^{q_1-q_2}&\ldots & \mu^{-1} \\
  1-\widetilde{H} & p_2 &\ldots & \vdots \\
  \vdots & \ddots & \ddots&\vdots\\
  \mu\,e^{q_n-q_1}&\ldots&1&p_n
\end{array}\right)\,.\label{ltr}
\eq
The corresponding $2\times 2$ matrix looks like
\bq
\widetilde{T}^{(1\ldots n)}=T^{(1\ldots n)}+
\left(\begin{array}{cc}
  H&0 \\
  0&0
\end{array}\right)\,T^{(3\ldots n)}
\,.\label{2ttr}
\eq
The separation variables $\{\l_1\,\l_2\,\ldots,\l_{n-1}\}$ for the
both system are zeroes of the  polynomials
\bq
{\mathbb C}(\l)=\delta\cdot {\cal L}_{12}(\l)=
\gamma\cdot \prod_{i=1}^{n-1}(\l-\l_i)\,,
\label{sepv}
\eq
where ${\mathbb C}(\l)$ be the entry of the matrix $T(\l)$ and
${\cal L}_{kj}(\l)$ be cofactor of the matrix $({\cal L}(\l)+\mu
I)$ associated with $(kj)$-entry. Additional set of variables is
defined by
\[\mu_i={\mathbb D}\,(\l_i)={\cal L}_{11}(\l_i)\,,\qquad i=1,\ldots, n-1\,\]
Notice that $\mu_i=\mu(\l_i)$ where $\mu(\l)$ is the eigenvalue of
$({\cal L}^{(n-1)}+\l I)$ and $T(\l)$.

According to (\ref{ltr}) and (\ref{2ttr}), namely these cofactors
of ${\cal L}_{11}(\l)$, ${\cal L}_{12}(\l)$ and these entries
${\mathbb C}(\l)$, ${\mathbb D}(\l)$ are invariant under change of
the time (\ref{trtan}). Thus, the variables $\{\l_i\,,\log
\mu_i\}$ are canonically conjugated which can be shown following \cite{skl85a}
by using (\ref{sklbr})
\[\{\l_i,\log \mu_k\}=\delta_{ik}\,.\]
From $\det T(\l)=1$ and $\det \widetilde{T}(\l)=(1-\widetilde{H})$
one immediately gets
\[
\begin{array}{lr}
  {\mathbb A}(\l_i)=\mu_i^{-1} &\qquad \mu_i+\mu_i^{-1}=P(\l_i)\,, \\
  \\
  \widetilde{\mathbb
A}(\l_i)=(1-\widetilde{H})\mu_i^{-1} &\qquad
\mu_i+(1-\widetilde{H})\mu_i^{-1}=\widetilde{P}(\l_i)\,.
\end{array}
\]
The original symplectic form is written as
\[
\Omega=\sum_{i=1}^{n-1} d\,\log \mu_i\wedge d\l_i + d \log\gamma\wedge d
p
\]
where $\gamma$ is defined by (\ref{sepv}). By using the standard
form of the hyperelliptic curves $C$ and $\widetilde{C}$
(\ref{todac}) and by applying Arnold's method \cite{arn89}, action
variables have the form
\ben
s_i&=&\oint_{\alpha_i} \dfrac12\left(\,
P(\l)+\sqrt{P(\l)^2-4\,}\right)\,d\l\,,\nn\\
\label{actoda}\\
\widetilde{s}_i&=&\oint_{\widetilde{\alpha}_i}
\dfrac12\left(\,
\widetilde{P}(\l)+\sqrt{\widetilde{P}(\l)^2-4(1-\widetilde{H})\,}\right)
\,d\l\,,\nn
\en
where $\alpha_i$ and $\widetilde{\alpha}$ are $\alpha$-cycles of the
Jacobi variety of the algebraic curves (\ref{todac}), respectively.
In fact, polynomials $P(\l),~\widetilde{P}(\l)$ and $\alpha$-cycles
depend on the constants of motion, which are dropped in the
notations. Thus, the Abel transformation linearizes equations of
motion by using first kind abelian differentials on the corresponding
spectral curves.

Finally let us consider the B\"{a}cklund transformation $B_\nu$
for the Toda lattice \cite{ga83,pgaud92}. As is well known
\cite{fml76}, transformation $B_\nu$ is canonical transformation
$(p,q)\mapsto(P,Q)$ of the initial phase space $\cal M$ preserving
all the integrals of motion (see \cite{skl98a} for a more detailed
list of properties of $B_\nu$).

For the Toda lattice canonical transformation $B_\nu$ can be
described by the generating function \cite{ga83}
\bq
F_\nu(q\,|\,Q)=\sum_{i=1}^n\left(e^{q_i-Q_i}-
e^{Q_i-q_{i+1}}-\nu(q_i-Q_i)\,\right)\,,
\label{fbakl}
\eq
such that
\bq
p_i=\dfrac{\partial F}{\partial q_i}\,,\qquad P_i=-\dfrac{\partial
F}{\partial Q_i}\,.
\label{momenta}
\eq
The B\"{a}cklund transformation $B_\nu$ has to preserve the
spectral invariants of the Lax matrices ${\cal L}(\mu)$ and
$T(\l)$. As a consequence, to prove that $B_\nu$ preserves
integrals of motion
\bq
I_k(p,q)=I_k(P,Q)\,,\label{iint}
\eq
one verifies that $B_\nu$ preserves the spectrum of the Lax matrix
$L(\mu)$ (\ref{alax})
\bq
{\sf M}(\mu,q,Q)\,{\cal L}(\mu,p,q)={\cal L}(\mu,P,Q)\,{\sf
M}(\mu,q,Q)
\,,\label{spl}
\eq
where
\bq
{\sf M}(\mu,q,Q)=\sum_{i=1}^{n-1}\, e^{Q_i-q_{i+1}}\,E_{i+1,i}+
\mu\,e^{Q_n-q_1}\,E_{1,n}\,.\label{splM}
\eq
(see \cite{ms91} for a detailed account of the theory of the
B\"aklund transformation as gauge transformation).

Canonical transformation (\ref{trtan}) of the extended phase space
${\cal M}_E$ associated with arbitrary index $1\leq j\leq n$
induces the following shift of the generating function
\bq
\widetilde{F}_\nu(q\,|\,Q)=
{F}_\nu(q,Q) +\widetilde{H}\,e^{Q_j-q_{j+1}}= {F}_\l(q,Q) +\Delta
F\,.
\label{shift}
\eq
Note, here $\widetilde{H}$ be independent variable of the extended
phase space ${\cal M}_E$. It means that in (\ref{momenta}) all the
partial derivatives $H$ with respect to any other coordinates of
${\cal M}_E$ are equal to zero. In this case the equality
(\ref{spl}) and the matrix ${\sf M}(\mu,q,Q)$ (\ref{splM}) are
invariant with respect to the change of the time.

As above, the same B\"{a}klund transformations $B_\nu$
(\ref{fbakl}) and $\widetilde{B}_\nu$ (\ref{shift}) are
isospectral deformations of the corresponding  $2\times 2$ Lax
matrices $T(\l)$ and $\widetilde{T}(\l)$. For the Toda lattices
the intertwining relations are equal to
\[
M_i(\l,\nu)\,T_i(p,q)=T_i(P,Q)\,M_{i+1}(\l,\nu)\,,
\]
where
\[
M_i(\l,\nu)=\left(\begin{array}{cc}
 1 & e^{Q_{i-1}} \\
 -e^{-q_i} & \nu-\l-e^{ Q_{i-1}-q_i}
\end{array}\right)\,,
\]
The same relations may be used after change of the time at $i\neq
j,j+1$. One additional non-factorized relation is given by
\ben
M_j(\l,\nu)\,&&\left[T_j(p,q)
\left(\begin{array}{cc}
 \l+p_{j+1} &a_j\, e^{q_{j+1}} \\
 -e^{-q_{j+1}}\,(1+\widetilde{H})& 0
\end{array}\right)
\right]
=\nn\\
\nn\\
=&&\left[T_j(P,Q)\left(\begin{array}{cc}
 \l+P_{j+1} &a_j\, e^{Q_{j+1}} \\
 -e^{-Q_{j+1}}\,(1+\widetilde{H})& 0
\end{array}\right)
\right]\,M_{j+2}(\l,\nu)\,.\nn
\en
The characteristic properties of the new B\"{a}klund
transformation $\widetilde{B}_\nu$ are verified following
\cite{skl98a}. To prove the spectrality property we have to use
one non-factorized relation as well.

Recall, the correspondence between the kernel of the corresponding
quantum Baxter ${\mathbb Q}$-operator and the function
$F_\l(q\,|\,Q)$ is given by the semiclassical relation
\cite{pgaud92,skl98a}. Change of the time (\ref{trtan}) gives rise to
factorization of the $\mathbb Q$-operator in the semiclassical
limit
\[\widetilde{\mathbb Q}\sim
\exp(-i \widetilde{F}/\hbar)= {\mathbb Q}\cdot
\exp(-i \Delta F/\hbar)\,.\]
Note, that harmonic oscillator may be mapped into the Coulomb
model by using canonical change of the time \cite{pe91}. In
quantum mechanics, such duality of the corresponding eigenvalue
problems has been used by Fok, Schr\"odinger and many other. As an
example, in the Birman-Schwinger formalism we can estimate
spectrum of the one Hamiltonian $\widetilde{H}$ by using known
spectrum of the dual Hamiltonian $H$. So, it will be interesting
to study such duality in framework of the quantum ${\mathbb
Q}$-operator theory.

\section{The Henon-Heiles and Holt integrable systems.}
\setcounter{equation}{0}
The Holt system is defined by the Hamilton function
\bq
\widetilde{H}({\bf p_x,p_y},{\bf x,y})=\dfrac12\,(\,{\bf p_x}^2+{\bf p_y}^2\,)+a\,{\bf x}^{-2/3}\,\,
(\,\dfrac{3\,b}4\,{\bf x}^2+{\bf y}^2+c\,)\,.
\label{holt}
\eq
Only three integrable cases are known \cite{rgb89}
\bq
(i)~b=1\,,\qquad(ii)~b=6\,,\qquad(iii)~b=16\,,
\eq
while the remaining parameters $a$ and $c$ be an arbitrary constants.
After canonical change of variables
\[
{\bf x}=\dfrac23\,x^{3/2}\,,\quad {\bf p_x}=p_x\,\sqrt{\,x\,}\,,\quad
{\bf y}=-\dfrac1{2\,\sqrt{3\,a}}\,p_y\,,\quad {\bf
p_y}=2\,\sqrt{\,3\,a\,}\,y\,.
\]
and rescaling
\[
a\to 4\,\left(\,\dfrac32\,\right)^{1/3}\,a\,,\qquad {c}
\to\dfrac{c}{3\,a}
\]
the Hamilton function (\ref{holt}) becomes
\bq
\widetilde{H}(p_x,p_y,x,y)=\dfrac{\,p_x^2+p_y^2\,}{2\,x}+2\,a\,\,
(\,b\,x^2+3\,y^2\,)+\dfrac{2\,c}x\,.
\label{ham}
\eq
According to \cite{rgb89,ts98b}, further canonical transformation
of the extended phase space
\bq
d\widetilde{t}=x\,dt\,,\qquad \widetilde{H}\mapsto
H=x\,\widetilde{H}\,,\nn\\
\label{hht}\\
\eq
preserves integrability and maps the Holt system into the
Henon-Heiles system
\bq
H(p_x,p_y,x,y)=\dfrac{\,p_x^2+p_y^2\,}{2}+2\,a\,x\,(\,b\,x^2+3\,y^2\,)+2\,c\,.
\label{tham}
\eq
At $b=1$ and at $b=16$ the corresponding Lax matrices are $3\times 3$
matrices and the spectral curves are trigonal algebraic curves
\cite{fo91}. We shall consider these more complicated cases in
the forthcoming publication.

At $b=6$ the $2\times 2$ Lax matrices for the Henon-Heiles system is
equal to
\cite{fo91, ts98b}
\ben
{\cal L}(\l)=
\left(\begin{array}{cc}
  {\mathbb A} & {\mathbb B} \\
  {\mathbb C} & -{\mathbb A}
\end{array}\right)(\l)
&=&
\left(\begin{array}{cc}p_x/2
-p_y\,y/4\,\l &\l+x-y^2/4\,\l\\
\\
 p_y^2/4\,\l&
 -p_x/2-py\,y/4\,\l
\end{array}\right)\nn\\
\nn\\ \nn\\
&+&6\,a\,\Bigl[\,\l^2-x\,\l+x^2+y^2/4\,\Bigr]\,
\left(\begin{array}{cc}0&0\\
1&0\end{array}\right)\,,\nn\\
\nn\\
\ll{laxii}
\en
\[
{\cal A}(\l)=\left(\begin{array}{cc}0&1\\
\\
6\,a\,(\,\l-2\,x\,)&0\end{array}\right)\,.\qquad\qquad\qquad\qquad
\]
Change of the time (\ref{hht}) gives rise to the mapping of these Lax
matrices
\bq
\widetilde{\cal L}(\l)={\cal L}(\l)-\dfrac12\,\widetilde{H}\,
\left(\begin{array}{cc}0&0\\
1&0\end{array}\right)\,,\qquad
\widetilde{\cal A}(\l)=\dfrac1{x}\,{\cal A}(\l)\,.
\label{l2tr}
\eq
and the following transformations of the  corresponding hyperelliptic
spectral curves
\ben
&\mu^2=P(\l)&=6\,a\,\l^3+\dfrac12\,{H-c}+\dfrac{K}{\l}\,,\nn\\
\label{henc}\\
&\mu^2=\widetilde{P}(\l)&=6\,a\,\l^3-\dfrac12\,\widetilde{H}\,\l-\dfrac{c}2
+\dfrac{\widetilde{K}}{\l}\,.\nn
\en
The corresponding transformation of the $r$-matrix Poisson brackets
has been considered in \cite{ts98b}.

The separation variables $\{\l_1\,\l_2\}$ for the both system are
zeroes of the  polynomial (see references in \cite{ts98b})
\bq
{\mathbb B}(\l)= \dfrac{(\l-\l_1)(\l-\l_2)}{\l}
\label{sepv2}
\eq
and
\[\mu_i={\mathbb A}\,(\l_i)\,,\qquad i=1,2\,.\]
Notice that $\mu_i=\mu(\l_i)$ where $\mu(\l)$ is the eigenvalue of
${\cal L}(\l)$.

According to (\ref{l2tr}), namely these entries of $T(\l)$ are
invariant under change of the time (\ref{hht}). These variables
are the standard parabolic coordinates, which lie on the
hyperelliptic curves (\ref{henc}), respectively. Applying Arnold's
method
\cite{arn89}, action variables have the form
\bq
s_i=\oint_{\alpha_i} \sqrt{P(\l)\,}\,d\l\,,\qquad
\label{actoda2}\\
\widetilde{s}_i=\oint_{\widetilde{\alpha}_i}
\sqrt{\widetilde{P}\,}\,d\l\,,\nn
\eq
where $\alpha_i$ and $\widetilde{\alpha}$ are $\alpha$-cycles of
the Jacobi variety of the algebraic curves (\ref{henc}),
respectively. Thus, the Abel transformation linearizes equations
of motion by using first kind abelian differentials on the
corresponding hyperelliptic spectral curves.

Change of the time (\ref{hht}) is related to ambiguity of the
corresponding Abel map \cite{ts98b}. In fact, these integrable
systems may be associated with the two different subsets of the
differentials into the complete basis of first kind abelian
differentials on the common hyperelliptic curve (\ref{henc}).

Now let us consider the known B\"{a}cklund transformation $B_\nu$
for the Henon-Heiles system \cite{w84}, which can be described by
the generating function
\ben
F(x,y\,|\,X,Y)&=&\sqrt{6\,a\,\nu}\,y\,Y +
\dfrac25\,\sqrt{6a\,(\nu-x-X)\,}\,\label{henF}\\
&\cdot&
\left( 2\l^{2} - (x +X)\l + 2x^{2} - x X + 2X^{2} +
\dfrac{5(\,y^{2}+Y^2)}{4}\,\right)\,.\nn
\en
The B\"{a}cklund transformation $B_\nu$ preserves the spectrum of
the Lax matrix ${\cal L}(\l)$ (\ref{laxii}) (for instance see
\cite{ms91,hkr99})
\bq
{\sf M}(\l,\nu){\cal L}(\l,x,y)={\cal L}(\l,X,Y){\sf M}(\l,\nu)
\,,\label{splhen}
\eq
where
\bq
{\sf M}(\l,\nu)=
\left(\begin{array}{cc}
 -\sqrt{6a\,(\nu-x-X)\,} & 1 \\
  6a\,(\l-x-X) & \sqrt{6a\,(\nu-x-X)\,}
\end{array}\right)
\,.\label{henM}
\eq
Change of the time (\ref{hht}) gives rise to the shift of the
generating function (\ref{henF})
\[
\widetilde{F}=F-\sqrt{6a\,(\nu-x-X)\,}\,\widetilde{H}
\]
and factorization of the corresponding kernel of ${\mathbb
Q}$-operator in the semiclassical limit.

Here the Hamiltonian $\widetilde{H}$ be independent variable of
the extended phase space ${\cal M}_E$. As for the Toda lattice,
equality (\ref{splhen}) and matrix $M(\mu,\nu)$ (\ref{henM}) are
invariant under change of the time.

\end{document}